\documentclass[aps,prl,a4paper,twocolumn,superscriptaddress,floatfix,showpacs]{revtex4}

\usepackage{graphicx}
\usepackage{dcolumn}
\usepackage{amsmath}
\usepackage{bm}

\begin{document}

\title{Matter-wave interferometry in a double well on an atom chip}

\author{T.~Schumm}
\affiliation{Physikalisches Institut, Universit\"at Heidelberg,
D-69120 Heidelberg, Germany} \affiliation{Laboratoire Charles
Fabry de l'Insitut d'Optique, UMR 8105 du CNRS, F-91403 Orsay,
France}
\author{S.~Hofferberth}
\author{L.~M.~Andersson}
\author{S.~Wildermuth}
\affiliation{Physikalisches Institut, Universit\"at Heidelberg,
D-69120 Heidelberg, Germany}
\author{S.~Groth}
\affiliation{Physikalisches Institut, Universit\"at Heidelberg,
D-69120 Heidelberg, Germany} \affiliation{Department of Condensed
Matter Physics, The Weizmann Institute of Science, Rehovot 76100,
Israel}
\author{I.~Bar-Joseph}
\affiliation{Department of Condensed Matter Physics, The Weizmann
Institute of Science, Rehovot 76100, Israel}
\author{J.~Schmiedmayer}
\email{schmiedmayer@atomchip.org} \affiliation{Physikalisches
Institut, Universit\"at Heidelberg, D-69120 Heidelberg, Germany}
\author{P.~Kr\"uger}
\email{krueger@physi.uni-heidelberg.de}
\homepage{http://www.atomchip.net} \affiliation{Physikalisches
Institut, Universit\"at Heidelberg, D-69120 Heidelberg, Germany}
\affiliation{Current address: Laboratoire Kastler Brossel,
\'{E}cole Normale Sup\'{e}rieure, F-75005 Paris, France}

\date{\today}

\begin{abstract}
Matter-wave interference experiments enable us to study matter at
its most basic, quantum level and form the basis of high-precision
sensors for applications such as inertial and gravitational field
sensing. Success in both of these pursuits requires the
development of atom-optical elements that can manipulate matter
waves at the same time as preserving their coherence and phase.
Here, we present an integrated interferometer based on a simple,
coherent matter-wave beam splitter constructed on an atom chip.
Through the use of radio-frequency-induced adiabatic double-well
potentials, we demonstrate the splitting of Bose-Einstein
condensates into two clouds separated by distances ranging from 3
to 80~$\mu$m, enabling access to both tunnelling and isolated
regimes. Moreover, by analysing the interference patterns formed
by combining two clouds of ultracold atoms originating from a
single condensate, we measure the deterministic phase evolution
throughout the splitting process. We show that we can control the
relative phase between the two fully separated samples and that
our beam splitter is phase-preserving.
\end{abstract}


\maketitle

The wave nature of matter becomes visible in interference
experiments \cite{Bad88}. Interferometry with atoms has become an
important tool for both fundamental and applied experiments in the
diverse fields of atomic physics, quantum optics, and metrology
\cite{Ber97}. For example, highly accurate acceleration
measurements have been based on atom interference \cite{Kas91}.
Even though in most cases, interference is fundamentally a single
particle phenomenon, the experiments become even more powerful
when performed with Bose-Einstein condensates (BECs).
Interferometry was used to demonstrate the remarkable property of
atoms in a BEC to possess a common phase \cite{And97}. Realizing
miniaturized matter-wave interferometers, as well as controlling
and engineering quantum states on a microscale in general, have
been long standing goals. Microscopic integrated matter-wave
devices \cite{Fol02} can be used to study the physics of
correlated many-body quantum systems and are promising candidates
for the implementation of scalable quantum information processing
\cite{Cir05}.

The combination of well-established tools for atom cooling and
manipulation with state-of-the-art microfabrication technology has
led to the development of atom chips
\cite{Fol02,Mue99,Rei99,Fol00,Dek00}. These devices have been
shown to be capable of trapping and guiding ultracold atoms on a
microscale. A variety of complex manipulation potentials have been
formed using magnetic \cite{Bru05}, electric \cite{Kru03}, and
optical fields \cite{Dum02}. BECs can be created efficiently in
such microtraps and coherent quantum phenomena such as
internal-state Rabi oscillations \cite{Tre04} and coherent
splitting in momentum space \cite{Wan05,Gue05} have been observed.

It is of particular interest to control the external (motional)
degrees of freedom of trapped atoms and spatially delocalized wave
packets on a quantum level \cite{Cal00a,Cha02}. A generic
configuration for studies of matter-wave dynamics is the double
well \cite{Shi04b,Alb05}. Dynamically splitting a single trap into
a double well is analogous to a beam splitter in optics and hence
forms a basic element of a matter-wave interferometer.
Interferometers on a microchip can be used as highly sensitive
devices because they allow measurements of quantum phases. This
enables experiments exploring the intrinsic phase dynamics in
complex interacting quantum systems (for example, Josephson
oscillations \cite{Alb05,Jos62}) or the influence of the coupling
to an external 'environment' (decoherence \cite{Zur03}).
Technologically, chip-based atom interferometers promise to be
very useful as inertial sensors on a microscale \cite{Kas02}. For
all these applications it is imperative that the deterministic
coherent quantum evolution of the matter waves is not perturbed by
the splitting process itself. Although several atom chip
beam-splitter configurations have been proposed and experimentally
demonstrated \cite{Kru03,Cas00,Mue00,Hom05,Shi05} none of them has
fulfilled this crucial requirement.

We present an easily implementable scheme for a phase preserving
matter-wave beam splitter. We demonstrate experimentally, for the
first time, coherent spatial splitting and subsequent stable
interference of matter waves on an atom chip. Our scheme is
exclusively based on a combination of static and radio-frequency
(RF) magnetic fields forming an adiabatic potential \cite{Mus87}.
The atomic system in the combined magnetic fields can be described
by a hamiltonian with uncoupled adiabatic eigenstates, so-called
dressed states. For sufficiently strong amplitudes of the RF
field, transitions between the adiabatic levels are inhibited as
the strong coupling induces large repulsion of these levels. This
property of the combined static and RF fields could be exploited
to form trapping geometries using the effective potential acting
on the dressed eigenstates \cite{Zob01}, and related demonstration
experiments with thermal atoms have been performed \cite{Col04}.

In a more general case than considered in ref. \cite{Zob01}, the
orientation, in addition to the intensity and frequency, of the RF
field determine the effective adiabatic potential V$_{\rm eff}$ at
the position {\bf r}:

\begin{eqnarray}
    \lefteqn{V_\text{eff} (\textbf{r}) =}&&\\
    &&m_{\rm F}\sqrt{[\mu_{\rm B}g_{\rm F}B_{\text{d.c.}}(\textbf{r}) - \hbar\omega_{\rm RF}]^{2} + [\mu_{\rm B}g_{\rm F}B_{\text{RF}\perp}(\textbf{r}) /2]^{2}}\nonumber
\end{eqnarray}

Here $m_{\rm F}$ is the magnetic quantum number of the state,
$g_{\rm F}$ is the Land\'{e} factor, $\mu_{\rm B}$ is the Bohr
magneton, $\hbar$ is the reduced Planck's constant, $B_{\rm d.c.}$
is the magnitude of the static trapping field and $\omega_{\rm
RF}$ is the frequency of the RF field. $B_{\rm RF\bot}$ is the
amplitude of the component of the RF field perpendicular to the
local direction of the static trapping field. The directional
dependence of this term implies the relevance of the vector
properties of the RF field, which enables the formation of a true
double-well potential.

\begin{figure}
    \includegraphics[width=\columnwidth]{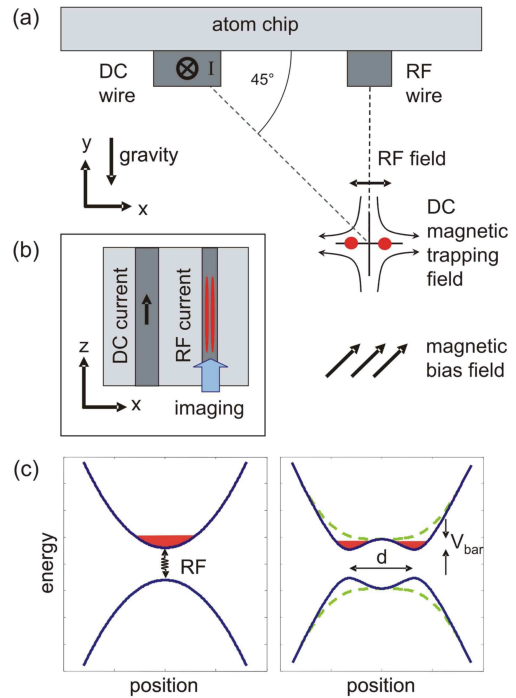}
    \caption{\label{fig1} Operation principle of the beam
    splitter. (a), A straight wire carrying a static (d.c.) current
    ($\sim$1~A) is used to trap a BEC on an atom chip directly below a
    second wire carrying a RF current ($\sim$60~mA at 500~kHz). The
    d.c. wire has a width of 50~$\mu$m, it is separated by 80~$\mu$m
    from the RF wire (width 10~$\mu$m). Placing the trap 80~$\mu$m
    from the chip surface at the indicated position allows for
    symmetric horizontal splitting. (b), Top view onto the atom chip
    (mounted upside down in the experiment): an elongated BEC is
    transversely split. All images are taken along the indicated
    direction. (c), Left: The RF magnetic field couples different
    atomic spin states (only two shown for simplicity). Right: the
    initial d.c. trapping potential is deformed to an effective
    adiabatic potential under the influence of the RF field with a
    frequency below the Larmor frequency at the trap
    minimum ($\sim$1~G). In the vertical ({\it y}) direction, the spatially
    homogeneous RF coupling strength leads to a slight relaxation of
    the static trap (dashed green line). Along the horizontal ({\it x})
    direction, the additional effect of local variations of the RF
    coupling breaks the rotational symmetry of the trap and allows for
    the formation of a double-well potential with a well separation
    (d) and potential barrier height $V_{\rm bar}$ (solid blue line). }
\end{figure}

Figure \ref{fig1} illustrates the operation principle of the beam
splitter. A standard magnetic microtrap \cite{Fol02} is formed by
the combined fields of a current-carrying trapping wire and an
external bias field; a static magnetic field minimum forms where
atoms in low-field-seeking states can be trapped. An RF field
generated by an independent wire carrying an alternating current
couples internal atomic states with different magnetic moments.
Owing to the strong confinement in a microtrap, the angle between
the RF field and the local static magnetic field varies
significantly over short distances, resulting in a corresponding
local variation of the RF coupling strength. By slowly changing
the parameters of the RF current we smoothly change the adiabatic
potentials and transform a tight magnetic trap into a steep double
well, thereby dynamically splitting a BEC without exciting it. We
accurately control the splitting distance over a wide range. The
potential barrier between the two wells can be raised gradually
with high precision, thus enabling access to the tunnelling regime
\cite{Alb05} as well as to the regime of entirely isolated wells.

The beam splitter is fully integrated on the atom chip, as the
manipulating potentials are provided by current-carrying
microfabricated wires. The use of chip-wire structures allows one
to create sufficiently strong RF fields with only moderate
currents and permits precise control over the orientation of the
RF field. Note that in our configuration the magnetic near-field
part of the RF field completely dominates.

We complete the interferometer sequence and measure the relative
phase between the split BECs by recombining the clouds in
time-of-flight expansion. In our experiments we found an
interference pattern with a fixed phase as long as the two wells
are not completely separated. The phase distribution remains
non-random and its centre starts to evolve deterministically once
the wells are entirely separated so that tunnelling is fully
inhibited on all experimental timescales.

\begin{figure}
    \includegraphics[width=\columnwidth]{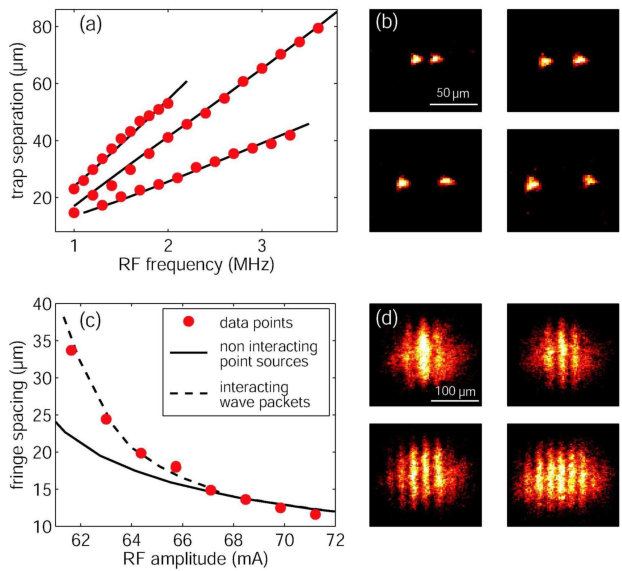}
    \caption{\label{fig2} The splitting of BECs is controlled over
    a wide spatial range. By adjusting amplitude and frequency of
    the RF field, we have been able to reach splitting
    distances of up to 80~$\mu$m. (a), A comparison of the measured splitting
    distances (red circles) to the theoretical expectation (black
    lines) yields good agreement for three different strengths of
    transverse confinement (gradients 1.1, 1.9 and 2.4~kG cm$^{-1}$, top to
    bottom). (b), The experimental data are derived from {\it in
    situ}
    absorption images (Roper Scientific MicroMAX: 1024BFT). (c), The
    fringe spacing is plotted as a function of RF amplitude (red
    circles). A simple approximation of the expected fringe spacing
    based on an expansion of a non-interacting gas from two points
    located at the two double-well minima agrees well with the data
    for sufficiently large splittings (solid line). For small
    splitting distances (large fringe spacings), inter-atomic
    interactions affect the expansion of the cloud. A numerical
    integration of the time-dependent Gross-Pitaevskii equation using
    our experimental parameters takes this effect into account (dotted
    line). (d), Interference patterns obtained after 14~ms
    potential-free time-of-flight expansion of the two BECs. For
    splittings below our imaging resolution ($d < 6\mu$m), the splitting
    distances can be derived from these interference patterns.}
\end{figure}

The experiments are performed in the following way. We routinely
prepare BECs of up to $\sim$10$^5$ rubidium-87 atoms in the F =
$m{_{\rm F}} =2$ hyperfine state in microtraps near the surface of
an atom chip \cite{Wil04}. Our smooth microwires \cite{Gro04}
enable us to create pure one-dimensional condensates (aspect ratio
$\sim$400) with chemical potential $\mu \sim \hbar \omega_{\perp}$
in a trap with high transverse confinement ($\omega_{\perp}=2 \pi
\times 2.1$~kHz) \cite{Kru05,Wil05c}. By tilting the external bias
field, we position the BECs directly below an auxiliary wire (Fig.
\ref{fig1}). A small sinusoidally alternating current through this
wire provides the RF field that splits the trap. For small
splitting distances ($<6~\mu$m) we ramp the amplitude of the RF
current from zero to its final value (typically 60-70~mA) at a
constant RF frequency ($\sim$500~kHz). This frequency is slightly
below the Larmor frequency of the atoms at the minimum of the
static trap ($\sim$1~G corresponding to $\sim$700~kHz). By
applying the ramp, we smoothly split a BEC confined in the
single-well trap into two. The splitting is performed transversely
to the long axis of the trap, as shown in Fig.~\ref{fig1}b. The
distance between the two wells can be further increased by raising
the frequency of the RF field (up to 4~MHz in our experiment). The
atoms are detected by resonant absorption imaging (see Fig.
\ref{fig2}) along the weak trapping direction, that is,
integrating over the long axis of the one-dimensional clouds. The
images are either taken {\it in situ} or after time-of-flight
expansion.

Unbalanced splitting can occur owing to the spatial inhomogeneity
of the RF field, owing to asymmetries in the static magnetic trap
and owing to gravity. Although the splitting process itself is
very robust, imbalances lead to a rapid evolution of the relative
phase of the two condensates once they are separated. The
influence of gravity can be eliminated by splitting the trap
horizontally. In the experiment we balance the double well by
fine-tuning the position of the original trap relative to the RF
wire.

To characterize the splitting, the split cloud is detected {\it in
situ}. We are able to split BECs over distances of up to 80~$\mu$m
without significant loss or heating (determined in time-of-flight
imaging). The measured splitting distances are in very good
agreement with the theoretical expectations for different
configurations of the initial single well (Fig. \ref{fig2}a).

\begin{figure}
    \includegraphics[width=\columnwidth]{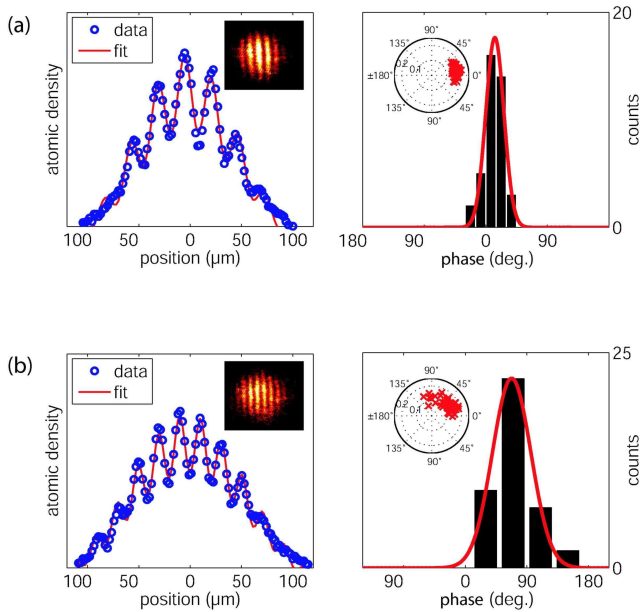}
    \caption{\label{fig3} The coherence of the splitting is
    examined by analysing matter-wave interference patterns. (a),
    Directly after (0.1~ms) the BECs have been split far enough
    ($d= 3.4~\mu$m) to inhibit tunnelling completely. (b), After (0.8~ms) the
    clouds have been taken farther apart ($d= 3.85 \mu$m). Left: a cosine
    function with a Gaussian envelope is fitted to the profiles
    derived from the two-dimensional images (insets). This yields
    information on fringe spacing, contrast and phase. Right: contrast
    and relative phase for 40 realizations of the same experiment are
    plotted in a polar diagram (inset). A histogram of the same data
    shows a very narrow distribution of the differential phase ($\sigma = 13^\circ$)
    directly after separating the clouds and a slightly broadened
    distribution ($\sigma = 28^\circ$) later in the splitting process. Both phase
    spreads are significantly smaller than what is expected for a
    random phase.}
\end{figure}

To study the coherence of the splitting process we recombine the
split clouds in time-of-flight expansion after a non-adiabatically
fast ($<50~\mu$s) extinction of the double-well potential. Typical
matter-wave interference patterns obtained by taking absorption
images 14~ms after releasing the clouds are depicted in Figure
\ref{fig2}d. The transverse density profile derived from these
images contains information on both the distance~{\it d} of the
BECs in the double-well potential and the relative phase~$\phi$ of
the two condensates. We determine the fringe spacing~$\Delta z$
and the phase~$\phi$ by fitting a cosine function with a Gaussian
envelope to the measured profiles (Fig. \ref{fig3}). For large
splittings ($d > 5~\mu$m for our experimental parameters), the
fringe spacing is given by $\Delta z$ = $ht/md$, where $h$ is
Planck's constant, $t$ is the expansion time and $m$ is the atomic
mass. This approximation of a non-interacting gas expanding from
two point sources is inaccurate for smaller splittings where the
repulsive interaction in the BEC has to be taken into account
\cite{Roe97}. Figure \ref{fig2}c shows the observed fringe spacing
that is compared with the above approximation and with a numerical
integration of the time-dependent Gross-Pitaevskii equation. We
find excellent agreement with the latter theory.

We assess the coherence properties of the beam splitter by
analysing the relative phase between the two condensates
throughout the splitting process. Figure \ref{fig3} shows the
results for 40 repetitions of the interference experiment
performed directly after the two condensates are separated (Fig.
3a) and after they have been taken farther apart (Fig.
\ref{fig3}b), respectively. We find a very narrow phase
distribution with a Gaussian width of $\sigma = 13^\circ$ and
$28^\circ$, respectively. Hence, the splitting process is
phase-preserving and the beam splitter is coherent.

\begin{figure}
    \includegraphics[width=\columnwidth]{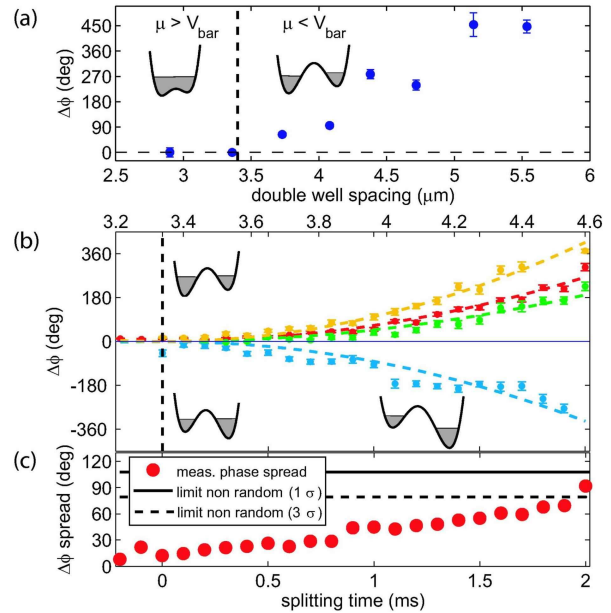}
    \caption{\label{fig4} Evolution of the differential phase
    throughout the splitting process. Error bars indicate the
    statistical error of the mean value. (a), Splitting the
    condensate to
    variable distances within 12~ms (splitting speed 1.4~$\mu$m ms$^{-1}$). The
    dashed vertical line indicates the trap separation for which the
    chemical potential $\mu$ of the BEC equals the potential barrier
    height $V_{\rm bar}~(3.4~\mu$m). As long as the barrier between the two wells
    is sufficiently low (left of the dashed line), the relative phase
    remains locked at zero. Once the wells are fully separated so that
    tunnelling is inhibited (right of the dashed line), the
    differential phase starts to evolve owing to a slight residual
    imbalance in the double well. (b), Relative phase of the two
    condensates throughout a slower splitting process (splitting speed
    0.6~$\mu$m ms$^{-1}$). The evolution of the differential phase is controlled
    by deliberately adjusting the double-well imbalance by displacing
    the trap. The observed evolution is in agreement with a numerical
    simulation based on our experimental parameters (dashed lines) for
    all data sets (yellow, red, green and blue points). Both signs of
    imbalance have been realized and the phase evolution is observed for
    a time (2~ms), which is more than four times the transverse
    oscillation period. (c), A typical distribution of the relative phase
    shows significantly non-random phases for the entire splitting
    process, the limits for a deviation by one and three standard
    deviations are indicated.}
\end{figure}

We have performed similar measurements throughout the entire
splitting process, starting from a well separation of $d \sim
3~\mu$m where the BECs are still connected to a splitting of $d
\sim 5.5~\mu$m. At larger $d$, the interference fringes are no
longer optically resolved. For splitting distances larger than
3.4~$\mu$m the potential barrier is sufficiently high to suppress
tunnel coupling between the two wells, so that the splitting
process is complete. For fast splitting we find the phase
distribution to be non-random over the whole splitting range (Fig.
\ref{fig4}a). In a more detailed experiment with slower splitting
we find that the spread is smaller than the expectation of a
randomized phase by more than three standard deviations for
splitting times shorter than 2~ms (Fig. \ref{fig4}c). Within these
limits, our data show an increase in phase spread and a coinciding
loss of average contrast. A possible explanation is the
longitudinal phase diffusion inside the individual one-dimensional
quasi-BECs \cite{Whi03}. This hypothesis is supported by the fact
that we always observe phase randomization (at finite interference
contrast) approximately 2.5~ms after the splitting is complete,
independent of the splitting distance $d$. This timescale roughly
agrees with the theoretical prediction for our experimental
parameters. More detailed experimental studies of the longitudinal
phase diffusion inside one-dimensional quasi-BECs are underway.

As the splitting is a coherent operation, we are able to measure
the phase evolution throughout the splitting process. We find that
the relative phase between the two condensates is locked to zero
as long as the chemical potential exceeds the potential barrier
($d < 3.4~\mu$m). Once the splitting is complete ($d > 3.4~\mu$m),
a deterministic phase evolution occurs (Fig. \ref{fig4}a,b). A
differential phase shift is induced by a slight residual imbalance
of the double-well potential (in the case shown in Fig.
\ref{fig4}a an energy difference of the order of $h \times 1$~kHz
$\mu$m$^{-1}$ additional splitting).

The double-well imbalance can be controlled by appropriately
adjusting the trap parameters. In our experiment we have studied
this by varying the current in the d.c. trapping wire and thereby
adjusting the position of the original single-well trap with
respect to the RF wire and the orientation of the splitting axis
with respect to gravity. The data depicted in Fig. \ref{fig4}b
show the resulting phase evolution for four different settings.
Again, the differential phase $\phi$ is zero as long as the clouds
are not fully separated; once tunnelling is inhibited, $\phi$
evolves quadratically with the split time. Although the
differential phase evolution in an asymmetric double well is
usually linear in time, in our case the wells are separated
further as the split time is increased, so that the imbalance
itself increases linearly with time. This leads to an overall
quadratic scaling, which is confirmed by a numerical integration
of the time-dependent Gross-Pitaevskii equation. We have varied
both the sign and strength of the imbalance in the double-well
potential. It is a crucial property of our beam splitter that the
balancing is fairly insensitive to changes of the controlling
parameter. The balancing can then be performed well above the
experimental noise level. In the illustrated case we have varied
the d.c. wire current on the per cent level ($\sim$10~mA
corresponding to a trap displacement of $\sim 1~\mu$m), whereas
the current stability is better than $10^{-4}$. In our experiment,
we have been able to balance the double well to an extent that
within one transverse oscillation period (0.5~ms) after the
condensates were fully separated, the phase shift remained smaller
than $18^\circ$.

There are several advantages of our atom chip beam-splitter
concept over previously implemented approaches: the splitting
distances are not limited by the structure size on the chip
\cite{Est05}, but rather by the ground-state size of the initial
single-well trap that can be orders of magnitude smaller. This
allows us to reach full splitting of a BEC of $1.1~\mu$m
transverse size (full width at half maximum) at a double-well
separation of only $3.4~\mu$m. The trapping wire, in contrast, has
a width of $50~\mu$m; the atoms are located at a distance of
$80~\mu$m from the surface. Furthermore, the dynamic splitting
process can be performed in a smooth (adiabatic) fashion, avoiding
excitations, by simply controlling the parameters of the RF field.

In conclusion, we have demonstrated coherent splitting and
interference of BECs using an atom chip. Our experiments are based
on a versatile microfabricated beam splitter that is fully
integrated on the chip. With our interferometer we have measured
and controlled the phase evolution between two BECs gradually
split over increasingly large distances. We are convinced that
such double-well potentials on atom chips are a starting point for
a variety of in-depth studies of the dynamics of one- and
three-dimensional quantum gases. Exploring the splitting and
recombination process in more detail is an immediate next step, as
are detailed studies of tunnelling and self-trapping in
low-dimensional quantum gases \cite{Alb05,Gio97}. Of particular
interest will be the time-dependent evolution and phase coherence
along low-dimensional correlated quantum systems in the fully
split and in the tunnelling regimes. Applications of our scheme
may range from investigations of atom-surface interactions and the
fundamental question of surface-induced decoherence to microscopic
atom interferometers for precision metrology. Last but not least,
simple and robust atom chip beam splitters and interferometers
based on our beam splitter may constitute the building blocks for
quantum information processing on the atom chip
\cite{Cir05,Cal00a}.

We thank H. Perrin and I. Lesanovsky for useful discussions. We
acknowledge financial support from the European Union, contract
numbers IST-2001-38863 (ACQP), MRTN-CT-2003-505032 (Atom Chips),
HPRN-CT-2002-00304 (FASTNet), HPMF-CT-2002-02022, and
HPRI-CT-1999-00114 (LSF) and the Deutsche Forschungsgemeinschaft,
contract number SCHM 1599/1-1.

\end{document}